# A Compact TIA in 22nm FDSOI CMOS for Qubit Readout in Monolithic Quantum Processors

Domenico Zito, AGH University of Science and Technology, 30-059 Krakow, Poland; domenico.zito@agh.edu.pl
Tan Doan Nhut, Aarhus University, 8200 Aarhus, Denmark

*Abstract*—This paper reports an inductorless transimpedance amplifier (TIA) with very compact size and adequate performance for spin qubit readout operations in monolithic quantum processors. The TIA has been designed and fabricated in a 22nm FDSOI CMOS foundry technology commercially available. The measurement results show a transimpedance gain of 103 dBΩ with a bandwidth of 13 GHz, at room temperature, and it is expected to exhibit slightly superior performance at cryogenic temperatures. The power consumption amounts to 4.1 mW. The core area amount to 0.00025 mm$^2$, i.e., about two orders of magnitude smaller with respect to the prior-art works, and approaching the qubit size, which makes the inductorless TIA a compact enabling solution for monolithic quantum processors.

*Keywords—Transimpedance amplifier, qubits, inductorless.*

## I. INTRODUCTION

One of the grand challenges toward the physical implementation of monolithic quantum processors is the development of reliable hardware technology platform capable of integrating millions of identical qubits together with the control and readout integrated circuits, all onto the same chip [1-4]. Regardless of the qubit nature (spin qubits, superconductive qubits, charge qubits, etc.), most of the current approaches are based on separate chips, with qubit array chip operating at extreme cryogenic temperatures (from a few tens to a few hundreds of millikelvins) packaged separately from the control and readout integrated circuit chips operating at a few kelvins (1-4 K), e.g., [5-10]. Although these approaches have concurred significantly to open new research perspectives toward the actual implementation of quantum processors, the multi-chip approach is heavily constrained by the connectivity issues, resulting in quite limited opportunities to fully address the grand challenge of the scalability of quantum processors to thousands and even millions of physical qubits. Moreover, because of the thermal operation regime at extreme cryogenic temperatures, the reduced dimensions and cooling capability of the cryostats limited to a few Watt [11], the power consumption and chip size requirements introduce further severe roadblocks to the actual scalability opportunities to large numbers of qubits, which add to the connectivity problem.

In such a challenging scenario of the research frontier on quantum technologies, the monolithic integration of qubits together with control and readout circuits onto a single chip opens to new further opportunities to address the quantum processors scalable to large numbers of qubits. In addition to the more compact and scalable implementations, eliminating the large chip-to-chip parasitics associated with the external circuitry alleviates significantly the connectivity problem and enables also the control and readout with lower noise, lower power and higher speed. Therefore, the opportunity to take advantage of the quantum effects in ultra-scaled devices together with the advanced circuit design capabilities emerging from the latest commercial complementary metal oxide semiconductor (CMOS) foundry technologies in mass production, offer concrete opportunities toward the implementation of scalable monolithic quantum processors, so accelerating the quantum leap into the quantum era. This perspective, as envisaged in [1-4], includes also the opportunity to move the qubits to higher cryogenic temperatures, so resulting less constrained to all the other aforementioned limitations. However, as anticipated in [3] new circuits with very low noise, wide bandwidth, very low power consumption and small form factor (area) are demanded to reach the superior performance requested in order to fully address these major challenges for the actual implementation of scalable monolithic quantum processors.

Innovative compact circuit solutions for control and readout of qubits have been reported for some building blocks such as microwave/mm-wave switches [12], amplifiers [13], and transimpedance amplifiers (TIAs) [1, 2]. In particular, the work in [2] reports a five stage high-gain TIA suitable for the intended qubit readout, together with an effective design strategy that allows designing the circuit at room temperature by taking the advantage of the validated MOSFET models, and then securing the same bias current density of the MOSFETs while the circuit is operating at cryogenic temperature by adjusting the back gate voltage. Such a circuit adopts three spiral inductors which have a relevant impact on the area footprint on chip. As a consequence, the layout area sets a limit to the density of qubits, each expected to be integrated together with the own control-and-readout circuitry, as much as possible (i.e., likely limited by the cross-talk and self-heating effects), in the very near proximity with the qubit itself. The preliminary design of a more compact TIA circuit solution with two spiral inductors has been reported in [3]. In order to enable a higher scalability to a large number of qubits, here we report an inductorless TIA featuring a very compact area approaching the size of the double quantum dot (qubit device) [3]. The TIA has been designed in 22nm fully-depleted silicon-on-insulator (FDSOI) CMOS and does not make use of integrated inductors (i.e., inductorless), so resulting in a very compact layout with adequate performance expected for electron/hole spin qubit readout [3].

The paper is organized as follows. Section II presents the inductorless TIA. Section III reports the implementation and measurements. Finally, in Section IV, the conclusions are drawn.

This work was supported in part by the European Commission through the European H2020 FET OPEN project IQubits (GA N. 829005).

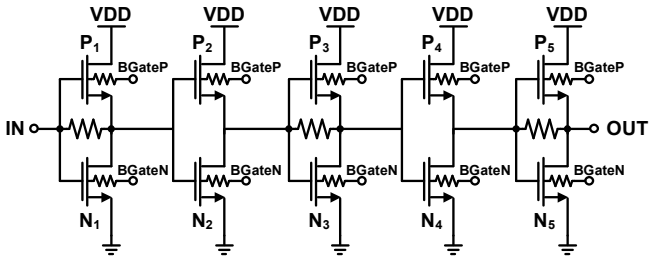

Fig. 1 Schematic of the five-stage inductorless TIA.

## II. INDUCTORLESS TIA

Fig. 1 shows the schematic of the five-stage inductorless TIA inspired from the previous designs [2, 3], but with no inductors.

The first stage is a CMOS inverting TIA stage with transistor sizes for a good tradeoff between gain, bandwidth and noise performances, whereas the following stages consist of two Cherry-Hooper amplifier stages with transistor sizes increased by a factor 2-3 (i.e., load driving capability, or fan-out) in order to maximize the gain-bandwidth product. The feedback resistor and transistors of the last stage are sized such that the output impedance of the circuit is matched to 50 Ω over the bandwidth of interest for the qubit readout operations. In this regard, as discussed in [3], it is worth mentioning that the very high-gain and impedance matching requirements are related to the envisaged qubit readout lab tests. In case of the prospected future monolithic integration, a gain of about 100 dBΩ and an impedance matching to the standard 50Ω load impedance are no longer required and therefore the TIA circuit design can be conveniently simplified and reduced to a lower number of stages [1], so to lead to a further reduction of the area footprint on chip and power consumption. Although the former is beneficial, but not critical due to the already very reduced area of the proposed solution approaching the qubit (i.e., double quantum dot) size, the latter is of paramount importance for further reducing the power consumption of the control and readout circuits per single qubit, which is largely predominant with respect the power consumption of the qubit itself, and therefore of strategic importance for the scalability to a large number of qubits.

Eliminating the inductors from the circuit produces the macroscopic evidence of a drastic reduction of the area footprint on chip. However, this visible and highly desired circuit feature has much deeper positive and enabling effects on the design of microwave and mm-wave building blocks, whose traditional design paradigm is based on cascaded active LC resonant stages made of transistors and large passive components, such as metal-oxide-metal (MoM) capacitors, spiral inductors and transformers, and transmission lines (TLs), e.g., [14-17]. As these are typically fabricated with the top thick metal layers of the back-end-of-line (BEOL) in order to limit the losses and capacitive effects toward the substrate, eliminating the inductors leads to eliminating their losses, as well as those due to their interconnections, including vias, from the transistor level all the way up to the top thick metal layers, so taking better advantage of the inherent performance of the transistors and avoiding the increase of current consumption needed to compensate for such losses [18-21]. Therefore, owing to the excellent transistor of the ultra-scaled technology nodes, eliminating the inductors may lead to innovative circuit design solutions with more compact area and lower power consumption, which are key for qubit control and readout in monolithic quantum processors scalable to large number of qubits, but also very attractive for microwave/mm-wave phase-array transceivers with a large number of array elements [21].

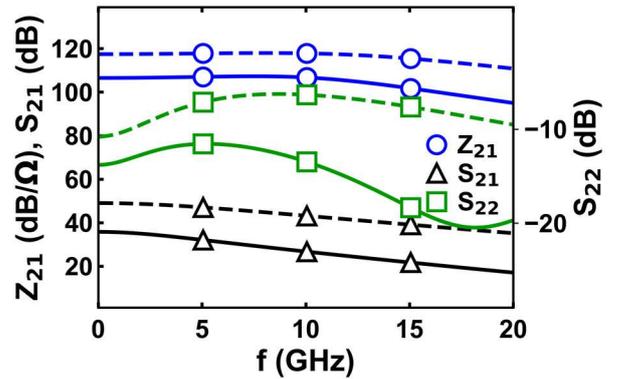

Fig. 2 Simulated $Z_{21}$, $S_{21}$, and $S_{22}$ parameters of the transimpedance amplifier at 300 K (solid lines) and 20 K (dashed lines).

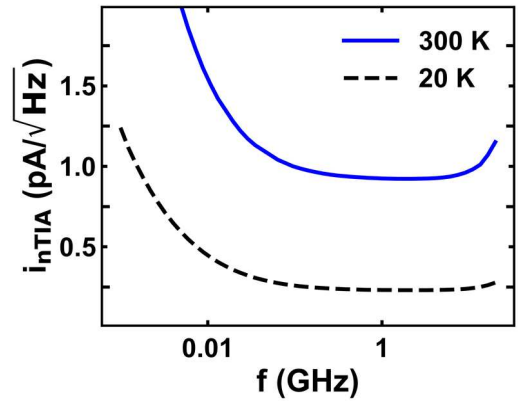

Fig. 3 Input-referred noise current spectral density of the transimpedance amplifier at 300 K and 20 K.

## III. DESIGN AND MEASUREMENTS

### A. Design implementation

In order to achieve the desired performance, the gate width of the n-MOSFET $N_1$ and the p-MOSFET $P_1$ have been sized to a width of 270 nm and 340 nm, respectively. For the impedance matching with the 50Ω output load impedance, the n-MOSFET and the p-MOSFET of the last stage of the TIA were sized with a width of 13.76 μm and 18.88 μm, respectively. All transistors have the minimum length offered by the 22nm FDSOI CMOS by GlobalFoundries.

Fig. 2 shows post-layout simulations (PLS) at 300 K and 20 K, for the typical bias voltages (VDD = 0.8 V, VBGateN = 1 V, VBGateP = -1.6V). The latter, is the lowest temperature for which convergence was achieved during the simulations. However, it is worth mentioning that the device models are not verified for such a low cryogenic temperature, thereby these results can be only considered as quite indicative, as the actual results will be available only from the measurements on the fabricated test-chips that are expected to be carried out in the next months in the cryogenic lab facilities available from the project partners [4]. In this paper we report the first experimental results achieved at room temperature. However, as already emerged from the previous building block designs, e.g., [2, 11, 12], we expect slightly superior performance at cryogenic temperatures as a consequence of the reduced resistive effects [1].

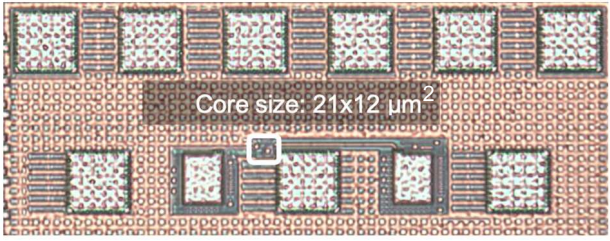

Fig. 4 Chip microphotograph of TIA with GSGSG and DC pads. The overall area of the test-chip amounts to 560 × 200 μm², including the pads. The core area in the white box amounts to 21 × 12 μm².

In particular, the simulations show that the TIA exhibits a $Z_{21}$ of 108.5 dBΩ and a -3dB bandwidth (BW) of 13 GHz, at room temperature. The output matching ($S_{22}$) is below -10 dB all over the bandwidth. The DC power consumption amounts to 4.5 mW. The simulation results at 20 K shows that slightly superior performances are expected at cryogenic temperatures. Despite the considerations above, the improvement of performance at cryogenic temperatures have been validated by cryogenic measurements at 2 K for all the pilot building blocks previously designed and tested [2, 11, 12], and therefore we expect slightly superior performance at cryogenic temperatures.

Fig. 3 shows the input-referred noise current density from post-layout simulations at 300 K and 20 K. The input noise current density is around 1 $pA/\sqrt{Hz}$ at 300K and reduces to 0.25 $pA/\sqrt{Hz}$ at 20K.

*B. Test-chip measurements*

The microphotograph of the TIA test-chip is reported in Fig. 4. The TIA has been measured on die with the Keysight Microwave Network Analyzer PNA-X N5245A. The measurement setup includes two FormFactor™ Infinity probes GSGSG i40 with a 100μm pitch, two 2.92mm coaxial cables and two 2.92-to-2.4mm adapters to probe the input and output pads of the TIA to the PNA-X. Fig. 5 shows the measurement setup.

Fig. 6 shows the measurement results at room temperature for the typical bias voltages. The TIA exhibits a measured transimpedance gain of 103 dBΩ with 13 GHz of bandwidth and $S_{22}$ around -10 dB all over the bandwidth. The DC power consumption amounts to 4.1 mW.

In order to measure the input equivalent current noise, the input of the TIA is left open circuited. Then, the output power of the TIA is measured. The output noise power is then converted into the equivalent input-referred noise by using the measured transimpedance gain. Fig. 7 shows that the TIA exhibits an input-referred noise current approximatively lower than 1.5 $pA/\sqrt{Hz}$ up to 13 GHz with a minimum noise current close to 1.1 $pA/\sqrt{Hz}$, in a good agreement with the simulation results.

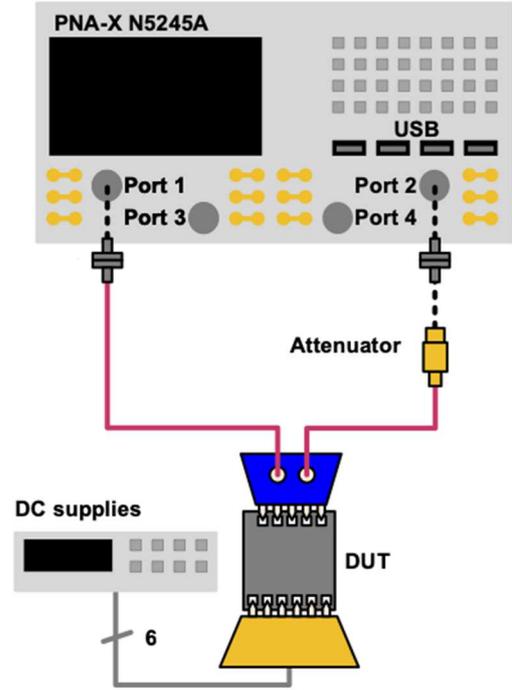

Fig. 5 Measurement setup.

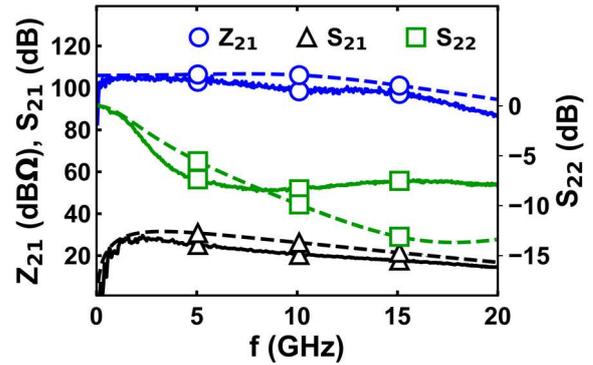

Fig. 6 Measured results (solid lines) of the TIA. Simulation results (dashed lines) are reported for a direct comparison with the measurements for the typical bias voltages.

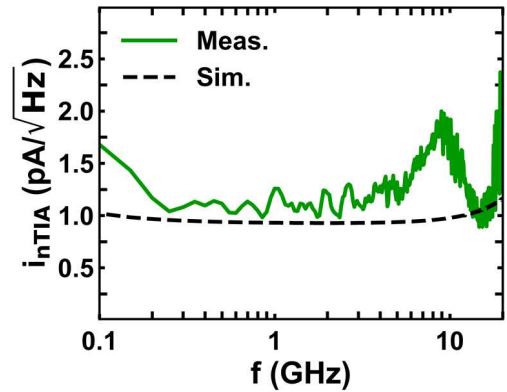

Fig. 7 Measured input-referred equivalent current noise (solid lines) of the TIA. Simulation results (dashed lines) are reported for a direct comparison with the measurements for the typical bias voltages.

Table I Comparison with prior-art TIAs.

| Ref. | This work | [1] | | [2] | | [22] | [23] |
|---|---|---|---|---|---|---|---|
| Tech. | 22nm FDSOI | 22nm FDSOI | | 22nm FDSOI | | 22nm FinFET | 16nm FinFET |
| Temp. (K) | 300 | 20 | 300 | 2 | 19 | 300 | 300 |
| BW (GHz) | 13 | 16[a] | 15 | 20 | 7.5 | 46.1 | 27 |
| $Z_{21}$ (dB) | 103 | 117[a] | 78 | 80 | 108 | 112 | 59.3 | 78 |
| $S_{22}$ (dB) | -10 | -8[a] | -10 | -10 | -15 | -15 | -10 | - |
| $P_{DC}$ (mW) | 4.1 | 4.6[a] | 3.1 | 3.1 | 4.5 | - | 11.2 | 107 |
| Input-ref. noise (pA/$\sqrt{Hz}$) | 1.1 | 0.25[a] | - | - | 0.82 | 0.19 | 12.6 | 16.7 |
| Area (mm$^2$) | 0.00025 | -- | | -- | | 0.025 | 0.02[b] |

[a] Indicative performance from simulations at 20 K. [b] Extrapolated from figure.

The measured input-referred $P_{1dB}$ of the TIA at 2.84 GHz (i.e., the peak gain frequency tone) amounts to −36.5 dBm.

Table I summarizes the performances of the inductorless TIA at room temperature (300 K) together with the indicative performances at cryogenic temperature (20 K), and together with the performances of the prior-art broadband CMOS TIAs. The inductorless TIA has a record compact area reduced by about two orders of magnitude, while exhibiting a high transimpedance gain comparable with the most representative prior-art works in ultra-scaled CMOS technologies.

IV. CONCLUSIONS

We have reported the design and implementation of a novel inductorless transimpedance amplifier in 22nm FDSOI CMOS with very compact size approaching the size of the double quantum dot, i.e., the qubit device, and with the desired gain and bandwidth requirements for qubit readout.

The transimpedance amplifier circuit was designed using a methodology aiming to eliminate all the inductors, so to minimize both the area and the losses due to the interconnections all through the back-end-of-line, from the transistor level all the way up to the top thick metal layers, so taking better advantage of the inherent performance of the transistors and reducing the power consumption. Also, the design methodology aimed to minimize the noise when driven by a double quantum-dot device, while achieving an adequate gain, bandwidth and broadband output matching to the 50Ω load impedance.

Overall, the results have demonstrated that the novel inductorless transimpedance amplifier circuit design reported here allows reducing significantly the footprint on chip by about two orders of magnitude with respect to microwave and mm-wave prior-art transimpedance amplifiers and reaching adequate measured performance aligned with the simulation results at room temperature, and expecting slightly superior performance at cryogenic temperatures.

ACKNOWLEDGMENTS

The authors are grateful to Keysight Technologies for their support through the donation of equipment and cad tools; Dr. C. Kretzschmar, Dr. P. Lengo, Dr. B. Chen (GlobalFoundries) for the technology support. The authors also grateful to Dr. Domenico Pepe (Renesas) for his contribution to the technical discussions.